\documentclass[12pt]{article}

\usepackage{scicite}
\usepackage{graphicx}
\usepackage{times}
\usepackage{journal_shortcuts}
\newenvironment{sciabstract}{%
\begin{quote} \bf}
{\end{quote}}

\newcounter{lastnote}
\newenvironment{scilastnote}{%
\setcounter{lastnote}{\value{enumiv}}%
\addtocounter{lastnote}{+1}%

\begin{list}%
{\arabic{lastnote}.}
{\setlength{\leftmargin}{.22in}}
{\setlength{\labelsep}{.5em}}}
{\end{list}}

\title{Baryons at the Edge of the X-ray Brightest Galaxy Cluster} 

\author
{Aurora Simionescu$^{1\ast}$, Steven W. Allen$^{1}$, Adam Mantz$^2$, Norbert Werner$^1$,\\ Yoh Takei$^3$, R. Glenn Morris$^1$, Andrew C. Fabian$^4$, Jeremy S. Sanders$^4$,\\ Paul E. J. Nulsen$^5$, Matthew R. George$^6$, Gregory B. Taylor$^{7\dagger}$\\
\\
\normalsize{$^{1}$KIPAC, Stanford University, 452 Lomita Mall, Stanford, CA 94305, USA}\\
\normalsize{$^2$ NASA Goddard Space Flight Center, Greenbelt, MD 20771, USA}\\
\normalsize{$^3$ Institute of Space and Astronautical Science (ISAS), JAXA, 3-1-1 Yoshinodai, Sagamihara,}\\
\normalsize{Kanagawa 229-8510, Japan}\\
\normalsize{$^4$ Institute of Astronomy, Madingley Road, Cambridge CB3 0HA, UK}\\
\normalsize{$^5$ Harvard-Smithsonian Center for Astrophysics, 60 Garden St., Cambridge, MA 02138, USA}\\
\normalsize{$^6$ Department of Astronomy, University of California, Berkeley, CA 94720, USA}\\
\normalsize{$^7$ University of New Mexico, Department of Physics and Astronomy, }\\
\normalsize{Alberquerque, NM 87131, USA}\\
\\
\normalsize{$^\ast$To whom correspondence should be addressed: asimi@stanford.edu.}\\
\normalsize{$^\dagger$Gregory B. Taylor is also an Adjunct Astronomer at the National Radio Astronomy Observatory.}
}

\date{}

\begin{document} 

\baselineskip24pt

\maketitle 

\begin{sciabstract}

Studies of the diffuse X-ray emitting gas in galaxy clusters have provided powerful constraints on cosmological parameters and insights into plasma astrophysics. However, measurements of the faint cluster outskirts have become possible only recently. 
Using data from the Suzaku X-ray telescope, we determined an accurate, spatially resolved census of the gas, metals, and dark matter out to the edge of the Perseus Cluster.
Contrary to previous results, our measurements of the cluster baryon fraction are consistent with the expected universal value at half of the virial radius.
The apparent baryon fraction exceeds the cosmic mean at larger radii, suggesting a clumpy distribution of the gas, which is important for understanding the ongoing growth of clusters from the surrounding cosmic web.

\end{sciabstract}

Galaxy clusters provide critical constraints on cosmological parameters which are independent from those determined using type Ia supernovae, galaxy surveys, and the primordial cosmic microwave background radiation (CMB) \cite{White93,vikhlinin2009,Mantz10a}. In particular, knowledge of their baryon content is a key ingredient in the use of clusters as cosmological probes \cite{Allen04,allen2008}. Most baryons reside in the hot, diffuse, X-ray emitting intra-cluster medium (ICM). Until recently, X-ray observations have generally targeted only the inner parts of clusters, where the emission is brightest, leaving a large fraction of their volumes practically unexplored. Estimates of the gas and total masses at large radii have relied on simple model extrapolations of the thermodynamic properties measured at smaller radii. 

X-ray spectroscopy of the outer regions of galaxy clusters was made possible only recently using the Suzaku satellite. Because of its much lower instrumental background than other X-ray observatories, Suzaku can measure the characteristics of the faint emission from cluster outskirts more reliably. Even so, few such observations have been published, and the thermodynamic profiles at large radii are not well resolved \cite{Reiprich09,George09,Bautz09,Hoshino10,Kawaharada10,Sato10}. 
The Perseus Cluster of galaxies is the brightest, extragalactic extended X-ray source. It is a relaxed system, both closer (at a redshift of 0.0183) and substantially brighter than any of the other clusters for which Suzaku has been previously used to study the ICM properties at large radii. Importantly, its large angular size mitigates the impact of residual systematic uncertainties in modeling the effects of Suzaku's complex point-spread function (PSF), making the Perseus Cluster an ideal target to study cluster outskirts.

A large mosaic of Suzaku observations of the Perseus Cluster, with a total exposure time of 260 kiloseconds, was obtained in August/September 2009. The pointings extend along two arms from near the cluster center towards the east (E) and northwest (NW), out to a radius of 2 degrees, which corresponds to 2.8 Mpc for a Hubble constant $H_0=70$ km/s/Mpc. Here, we focus on the data obtained with the three available X-ray imaging spectrometer (XIS) cameras (see supporting online text and Fig. 1).
We extracted spectra from annuli centered on the cluster center. After accounting for background emission, we modeled each spectral region as a single-temperature thermal plasma in collisional ionization equilibrium, with the temperature, metallicity, and spectrum normalization as free parameters. 

The best-fit radial profiles of temperature and metallicity are presented in Fig. 2. Individual elements are assumed to be present with Solar relative abundances\cite{Feldman92}. For comparison, we also show results previously obtained from an ultra-deep Chandra observation of the cluster center\cite{sanders2007}. The Suzaku and Chandra data sets show excellent agreement where they intersect, and together measure the temperature and metallicity structure of the intra-cluster gas with high precision and spatial resolution out to the virial radius (defined here as $r_{200}$, the radius within which the mean enclosed mass density of the cluster is 200 times the critical density of the Universe at the cluster redshift). In the narrow interval spanning 0.95-1.05$r_{200}$, the temperature is approximately a third of the peak temperature. Along the eastern arm, between 0.1 and 0.7 Mpc, the temperature is systematically lower than towards the northwest, and the X-ray emission is brighter. This thermodynamic feature is known as a ``cold front'' and typically arises following a merger between the main cluster and a smaller sub-cluster\cite{markevitch2007}. 

Our results show that the cluster outskirts are substantially metal-enriched, to a level amounting to approximately one third of the Solar metallicity. Previously, the only measurement of the metallicity close to the virial radius was obtained from a large region spanning 0.5--1 $r_{200}$ near the compressed outskirts of two interacting clusters\cite{fujita2008}. This previous result is in agreement with our measurements when converted to the Solar abundance units\cite{Feldman92} adopted here.

From the Suzaku data, we have also determined the electron density, entropy, and pressure profiles, corrected for projection effects under the assumption of spherical symmetry (Fig. 3). Outside the cold front at 0.7 Mpc, there is a good match between the E and NW profiles, with the electron density decreasing steadily with radius, approximately following a power-law model $n_e\propto r^{-\alpha}$ with slope $\alpha=1.68\pm0.04$. This is consistent with previous results from ROSAT data extending out to $\sim1.4$ Mpc\cite{Ettori98}.

Standard large scale structure formation models show that matter is shock heated as it falls into clusters under the pull of gravity. Simple theoretical models of this process predict that the entropy $K$ should behave as a power-law with radius, $K\propto r^{\beta}$, with $\beta\sim1.1 - 1.2$\cite{tozzi2001,voit2005MNRAS}. Except for the eastern cold front region, the entropy profile in Perseus roughly follows this expected trend until 2/3 $r_{200}$. Beyond this radius and until $0.95 r_{200}$, both the E and NW arms show a flattening from the power-law shape, confirming hints from previous Suzaku results\cite{George09}. 

The pressure profile is the most regular of the thermodynamic quantities plotted, and at most radii shows good agreement between the E and NW. At large radii, the pressure profile appears significantly shallower than expected by extrapolating the average profile of a sample of clusters studied previously with the XMM-Newton satellite within $\sim$0.5$r_{200}$\cite{arnaud2010}. 

Invoking hydrostatic equilibrium of the ICM, the gas pressure can be used to estimate the underlying gravitational potential and total (dark plus luminous matter) mass profile of the cluster. Numerical simulations show that the latter typically follows a functional form described by Navarro, Frenk \& White\cite{navarro1997}, also known as the NFW profile. We used the data from the northwestern arm of the Perseus Cluster, which appears dynamically relaxed, to determine the best-fit total mass profile, assuming an NFW form (supporting online text). The best-fit mass model parameters are typical of those predicted from numerical simulations; the NFW model provides a good description of the Suzaku data. 

Measuring the total mass profile allowed us to calculate the virial radius of the cluster, $r_{200}=1.79\pm0.04$ Mpc, the corresponding enclosed total mass $M_{200}=6.65^{+0.43}_{-0.46}\times10^{14}$ solar masses, and the cumulative gas mass and gas mass-to-total mass fraction, $f_{gas}$, as a function of radius (Fig. 4). 
At relatively small radii of 0.2--0.3$r_{200}$, the measured $f_{gas}$ value is in good agreement with direct measurements from the Chandra X-ray Observatory\cite{allen2008} and measurements of the Sunyaev-Zel'dovich (SZ) effect\cite{laroque2006} for two large samples of galaxy clusters. 
At about half of $r_{200}$, the integrated gas mass fraction reaches the cosmic mean value computed from the CMB\cite{komatsu2010}, considering that on average 12\% of the baryons are in stars\cite{lin2004,gonzalez2007} and the rest are in the hot X-ray emitting gas phase. 
Outside 2/3 of the virial radius, where the entropy also deviates from the expected power-law behavior, we find that the apparent $f_{gas}$ exceeds the cosmic mean baryon fraction measured from the CMB\cite{komatsu2010}. 

The most plausible explanation for this apparent excess of baryons at large radius is gas clumping. In the X-rays, the directly measurable quantity from the intensity of the bremsstrahlung emission is the average of the square of the electron density, $\left<n_e^2\right>$, rather than $\left<n_e\right>$. If the density is not uniform, i.e. the gas is clumpy, which is expected to occur as matter falls into the cluster, the average electron density estimated from the bremsstrahlung intensity will overestimate the true average, affecting the gas density, gas mass fraction, entropy, and pressure profiles. 

Outside the central region, and inside the radius where clumping becomes important, the measured $f_{gas}$ profile shows good agreement with recent numerical simulations\cite{young2010}, where a semi-analytic model was used to calculate the energy transferred to the intra-cluster gas by supernovae and AGN during the galaxy formation process. This model did not include a realistic implementation of gas cooling, and does not capture the complex processes in the central cool core of the cluster; the model is therefore not plotted in this region. Extrapolating this model into the outskirts where clumping is important, we used its predictions together with the measured $f_{gas}$ to determine by how much the electron density must be overestimated to produce the difference between the data and the model. This factor (plotted in green in the bottom panel of Fig. 4) reaches a value of up to 4 in the last annulus centered around the virial radius. The dense clumps are likely to be infalling and may be confined by ram pressure. 

Correcting the electron density using this factor, and accordingly the entropy and pressure profiles, we obtained the red lines shown in Fig. 3. The clumping-corrected entropy profile along the northwestern arm is consistent with the expected power-law profile. 
Moreover, the clumping-corrected pressure is also consistent with that expected by extrapolating the average profile of a sample of clusters previously studied with XMM-Newton\cite{arnaud2010}. The corrected electron density decreases more steeply with increasing radius, with the best-fit power-law index becoming $\alpha=2.5$. Correcting for clumping therefore seems to offer a simple solution to all the potential puzzles posed, at first glance, by the observed thermodynamic profiles of the Perseus Cluster at large radii. No additional physics is required by the data. 

Our study shows no evidence for the puzzling deficit of baryons at $r \ge 0.5 r_{200}$ inferred from some previous studies of other systems, using lower quality data and/or extrapolated models\cite{Ettori03,McCarthy07,afshordi2007,vikhlinin2006,Andreon10}. This suggests that within $r < 0.5 r_{200}$ the physics of the X-ray emitting gas is relatively simple and X-ray measurements can be used robustly for cosmological work. At larger radii, however, the cluster gas is significantly clumped. 

Numerical simulations predict gas clumping in the cluster outskirts \cite{Roncarelli06}. However, the amount of clumping depends on a large number of physical processes in the ICM which are currently uncertain, for example viscosity, conduction, star formation feedback, and magnetic fields. Although our results were obtained for just one galaxy cluster, it is expected that the observed physical processes are common. Our results therefore provide an anchor for numerical models of ICM physics and for simulations of the formation and ongoing growth of galaxy clusters. An independent measurement of gas clumping can be obtained from the combination of X-ray and SZ observations, which have different dependences on the electron density. 

\bibliography{bibliography,clustersnewest,clustersvirial}

\bibliographystyle{Science}

\begin{scilastnote}
\item We thank Peter Thomas and Owain Young for kindly providing the simulation results shown in Fig. 4. Support for this work was provided by NASA through Einstein Postdoctoral Fellowship grants number PF9-00070 and PF8-90056 awarded by the Chandra X-ray Center, which is operated by the Smithsonian Astrophysical Observatory for NASA under contract NAS8-03060. We acknowledge NASA grants NNX09AV64G and NNX10AR48G, issued through the Suzaku Guest Observer program, and NNX08AZ88G. The authors thank the Suzaku operation team and Guest Observer Facility, supported by JAXA and NASA. This work was supported in part by the U.S. Department of Energy under contract number DE-AC02-76SF00515. We also acknowledge the Grant-in-Aid for Scientific Research of MEXT Japan (KAKENHI No. 22111513) and Chandra award GO0-11138B.
\end{scilastnote}

\noindent{\bf Supporting Online Material}\\
www.sciencemag.org\\
Supporting text\\

\clearpage

\begin{figure*}
\begin{center}
\includegraphics[width=0.9\textwidth]{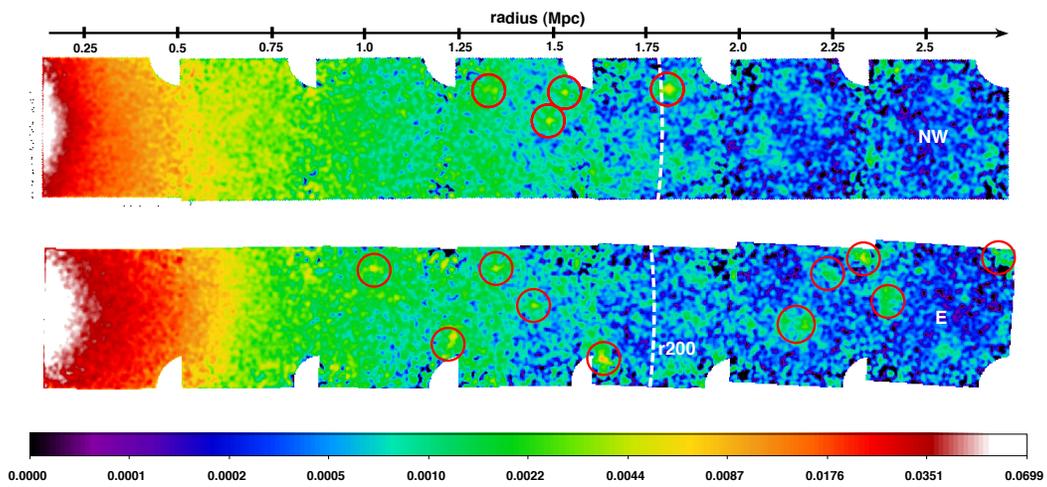}
\end{center}
\caption{\footnotesize 
X-ray surface brightness image of the NW (top) and E (bottom) arm mosaics observed with Suzaku, corrected for vignetting and instrumental background. The dashed white line marks the virial radius; the red circles mark excluded point sources and instrumental artefacts. The images have been rotated so that the cluster center is towards the left.
}
\label{sbimg}
\end{figure*}

\begin{figure}[p]
\includegraphics[width=0.9\columnwidth]{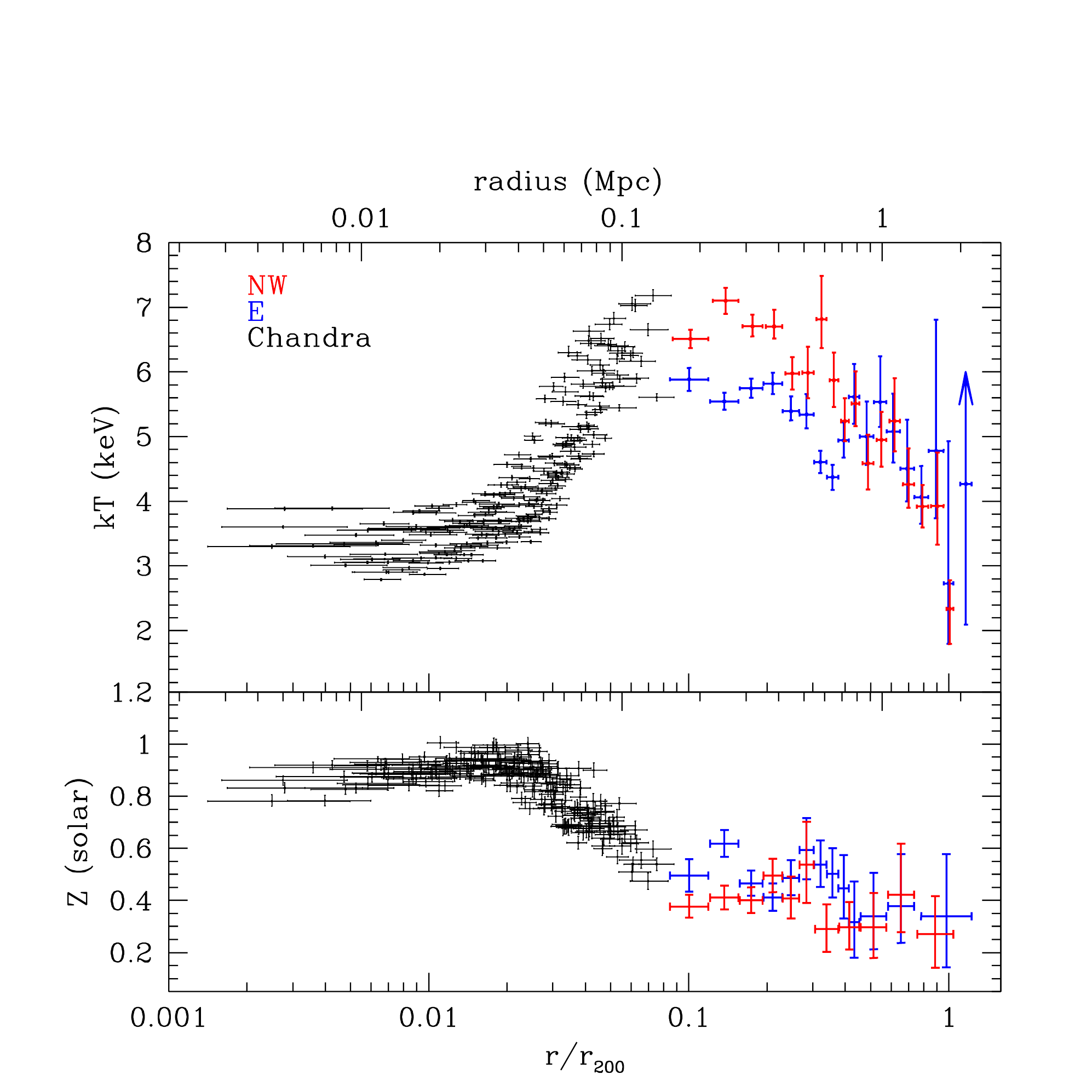}
\caption{\footnotesize 
Projected temperature (kT) and metallicity (Z) profiles of the Perseus Cluster. Results from Suzaku observations of the northwestern arm are shown in red, and of the eastern arm in blue. Earlier Chandra measurements of the cluster center\cite{sanders2007} are shown in black.
}
\label{tfe}
\end{figure}

\begin{figure}[p]
\includegraphics[width=0.8\columnwidth]{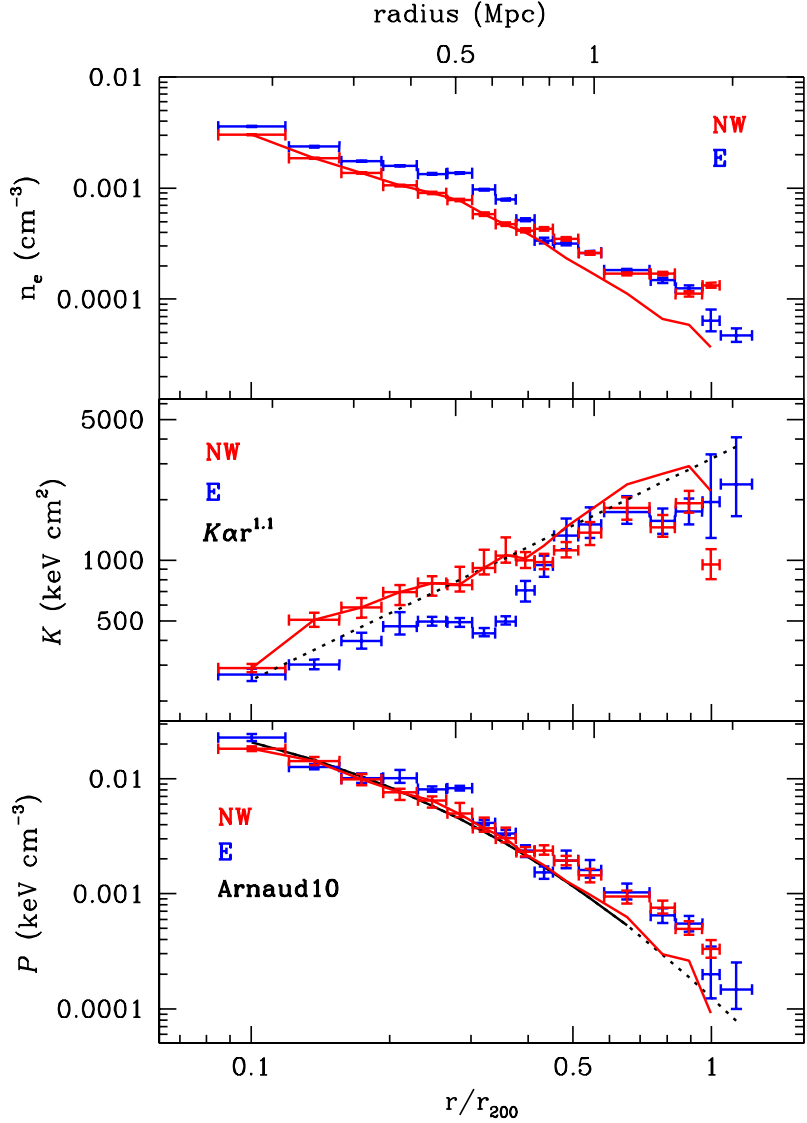}
\caption{\footnotesize 
Deprojected electron density ($n_e$), entropy ($K$), and pressure ($P$) profiles towards the northwest (red data points) and east (blue data points). The red line shows the northwestern profiles corrected for clumping. The expected entropy profile from simulations of gravitational collapse\cite{tozzi2001,voit2005MNRAS} is a power-law with index $\beta\sim1.1$, over-plotted as a black dotted line in the entropy panel. The average profile of a sample of clusters previously studied with the XMM-Newton satellite within $\sim$0.5$r_{200}$\cite{arnaud2010} is shown with a solid black curve in the pressure panel; its extrapolation to $r_{200}$ is shown with a dotted black line.
}
\label{nesp}
\end{figure}

\begin{figure}[p]
\includegraphics[width=0.9\columnwidth]{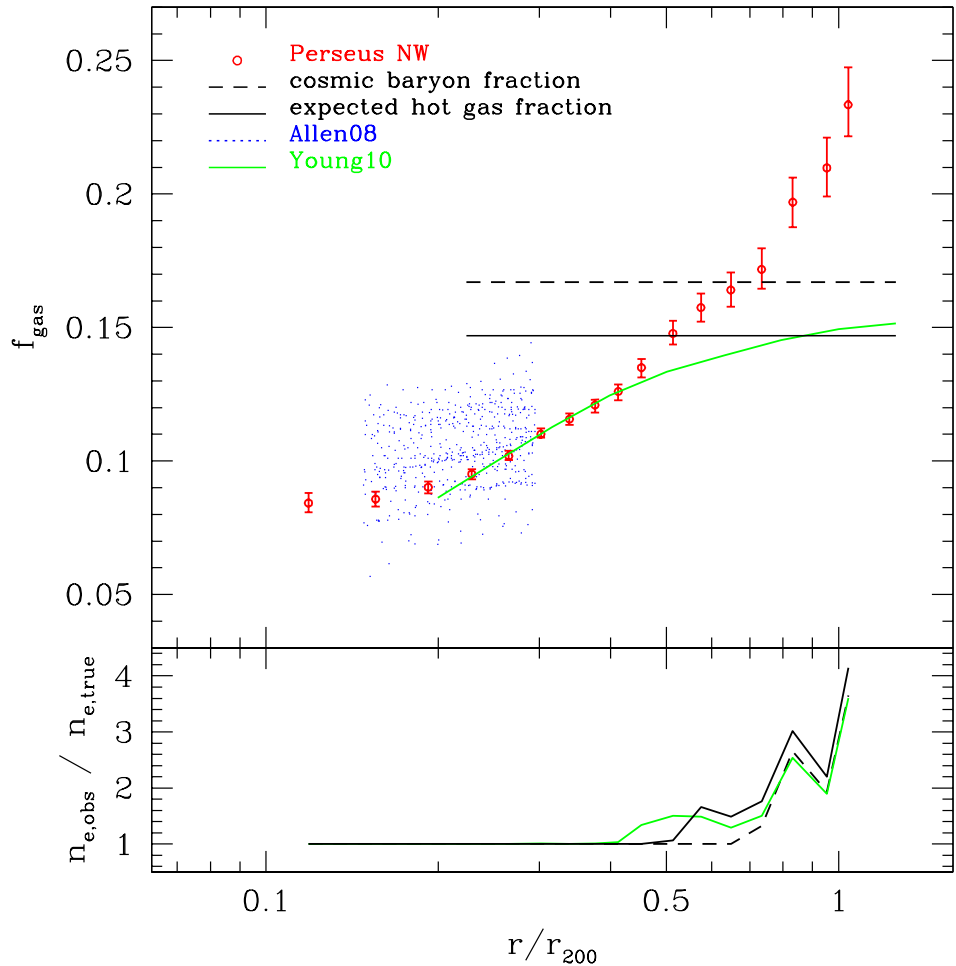}
\vspace{-1cm}
\caption{\footnotesize 
The integrated, enclosed gas mass fraction profile for the NW arm. The cosmic baryon fraction from WMAP7\cite{komatsu2010} is indicated by the horizontal dashed black line; accounting for 12\% of the baryons being in stars\cite{lin2004,gonzalez2007} gives the expected fraction of baryons in the hot gas phase, shown as a solid black line. The values previously measured for a sample of relaxed clusters at smaller radii with Chandra\cite{allen2008} are shown with blue dots. Predictions from numerical simulations\cite{young2010} are shown in green. The bottom panel shows by how much the electron density should be overestimated in each annulus due to clumping in order for the cumulative $f_{gas}$ not to exceed the correspondingly colored curves in the plot above.
}
\label{fgas}
\end{figure}


\begin{center}
{\bf\large Baryons at the Edge of the X-ray Brightest Galaxy Cluster}\\
{\bf Supporting text}
\end{center}

{\bf Observations and data analysis} 

Below, we describe the observations and data analysis methods in more detail. 
The Suzaku mosaic of the Perseus Cluster consists of seven pointings extending towards the east and seven towards the northwest of the cluster center. The nominal exposure times moving outward along each arm are 10, 10, 10, 20, 20, 30, and 30~kiloseconds (ks). 

The standard screening criteria recommended by the XIS instrument team\footnote{Arida, M., XIS Data Analysis, http://heasarc.gsfc.nasa.gov/docs/suzaku/analysis/abc/node9.html (2010)} were applied to create a set of cleaned event lists. We removed times of low geomagnetic cut-off rigidity (COR$\leq$ 6 GV) and verified that our observations were not affected by solar wind charge exchange (SWCX) emission using the method described in ({\it S1}). Images and spectra of the instrumental background were determined from dark Earth observations and subtracted from the data. The vignetting correction was performed using ray-tracing simulations of extended, spatially uniform emission ({\it S2}). 

We extracted and co-added images in the 0.7--7 keV energy band for each pointing using the three XIS detectors. Spectra were extracted from annuli centered on the cluster center, excluding the visually identified bright point sources and instrumental artefacts marked by red circles in Fig. 1 of the manuscript. We also excluded a $30^{\prime\prime}$ region around the detector edges and around the calibration sources at the detector corners. 

The spectra were binned to a minimum of one count per bin and fit in XSPEC ({\it S3}) version 12.6 using the extended C-statistic. We included in the spectral fitting data in the range 0.7--7 keV and where the predicted contamination by stray light from the bright cluster center was below 20\% of the best-fit cluster signal. For the Galactic absorption, we used the average column densities from the Leiden-Argentine-Bonn radio HI survey ({\it S4}). The cosmic X-ray background (CXB) was modeled with four components: three thermal models accounting for the local hot bubble, the Galactic halo, and a hot foreground present at low Galactic latitudes, and a power-law to account for the integrated emission of unresolved point sources. Data from the Suzaku mosaic beyond the estimated virial radius of Perseus, as well as from the ROSAT All Sky Survey (RASS), were used to constrain the parameters of the CXB model and estimate their spatial variations. Systematic uncertainties due to expected variations of the CXB are typically below the 1$\sigma$ statistical error in all the profiles shown in the manuscript. 
After accounting for background emission, each individual spectral region was modeled as single-temperature thermal plasma in collisional ionization equilibrium, using the {\it apec} code ({\it S5}). 

\bigskip

{\bf NFW mass model}

The Perseus Cluster is a relatively relaxed system, as indicated by its regular X-ray morphology and moderately strong cool core. There is no evidence for an ongoing major merger either from X-ray data or from the internal galaxy distribution. A minor merger is readily identified in the thermodynamic profiles along the E arm, but no similar signatures are seen toward the NW. The NW arm is thus a particularly relaxed direction in a cluster that is relatively relaxed overall.

To constrain the mass profile for the cluster, we used the Suzaku data for the NW arm and Chandra measurements for 4 inner annuli (0.02--0.08$r_{200}$), which capture the decrease in ICM temperature at small radii. Due to residual cross-calibration uncertainties ({\it S6}), the statistical precision of the Chandra measurements was limited not to exceed that of the Suzaku data, effectively using only a fraction of the available Chandra data.

The mass analysis employs the standard assumptions of hydrostatic equilibrium and spherical symmetry (the latter is a good approximation when modeling the X-ray emission from relaxed galaxy clusters; {\it S7}). We parametrized the mass distribution using an NFW model, which is well motivated by simulations of cosmological structure formation ({\it S8}). The cluster atmosphere was modeled as piecewise isothermal (constant temperature in each shell). Applying the hydrostatic equation to the set of temperatures in each annulus, along with the NFW mass model, specifies the density profile of the ICM up to an overall normalization factor. This non-parametric description of the ICM allows considerably more freedom than methods which require the temperature and density profiles to follow simple analytic functions. This is particularly important when modeling the outskirts of clusters, for which no robust prior expectations for the forms of the gas density and temperature profiles exist.

This model was implemented in XSPEC as an adaptation of the CLMASS mixing model introduced by ({\it S9}). The predicted model emission from each spherical shell is projected and fitted to the spectra for all annuli simultaneously. The fitting procedure used a Markov Chain Monte-Carlo algorithm to determine posterior distributions for the gas mass, total mass and gas mass fraction profiles. The error bars shown in the upper panel of Fig. 4 are the marginalized (one-dimensional) 68.3 percent confidence intervals for the integrated $f_{gas}$ values at the outer edge of each isothermal shell.

The minimum C-statistic (corresponding to the best-fitting parameters) is within the range expected from Monte Carlo simulations of the data, indicating that the model provides an acceptable fit. The best-fitting NFW function has a concentration parameter $c=5.0\pm0.5$, which falls comfortably within the expected range for clusters of the measured mass, as determined by numerical simulations. The measured mass also agrees well with expectations based on simple mass-temperature scaling relations (e.g. {\it S10}). The gas temperatures and densities from the fit are in excellent agreement with the values determined from simple spectral deprojection.

Our analysis shows that gas clumping becomes significant in the outer 5 annuli. Relaxing the assumption of hydrostatic equilibrium in these regions, while still accounting for the impact of projected emission appropriately, leads to consistent results on the gas mass fraction profile. Similarly, excluding the central 4 annuli (those using Chandra data) does not significantly affect the results; nor does relaxing the mass model to a generalized NFW form, which includes the central density gradient as an additional free parameter.


\bigskip

{\bf References}\\
\smallskip
S1.   R. Fujimoto, {\it et al.}, {\it PASJ} {\bf 59}, 133 (2007). \\
\smallskip
S2.   Y. Ishisaki, {\it et al.}, {\it PASJ} {\bf 59}, 113 (2007). \\
\smallskip
S3.   K. Arnaud, {\it ASPC} {\bf 101}, 17 (1996). \\
\smallskip
S4.   P. M. W. Kalberla, {\it et al.}, {\it A\&A} {\bf 440}, 775 (2005). \\
\smallskip
S5.   R. K. Smith, N. S. Brickhouse, D. A. Liedahl, J. C. Raymond, {\it ApJ} {\bf 556}, L91 (2001). \\
\smallskip
S6.   M. Tsujimoto, {\it et al.}, {\it A\&A} {\bf 525}, A25 (2011). \\
\smallskip
S7.   D. Nagai, A. Vikhlinin, A. V. Kravtsov, {\it ApJ} {\bf 655}, 98 (2007). \\
\smallskip
S8.   J. F. Navarro, C. S. Frenk, S. D. M. White, {\it ApJ} {\bf 490}, 493 (1997). \\
\smallskip
S9.   P. E. J. Nulsen, S. L. Powell, A. Vikhlinin, {\it ApJ} {\bf 722}, 55 (2010). \\
\smallskip
S10. M. Arnaud, E. Pointecouteau, G. W. Pratt, {\it A\&A} {\bf 441}, 893 (2005).\\

\end{document}